\documentclass[12pt]{iopart}

\usepackage{amssymb}
\usepackage{graphicx}
\eqnobysec

\usepackage{epsfig}

\begin{document}

\title{Uniform tiling with electrical resistors}

\author{J\'ozsef Cserti$^1$, G\'abor Sz\'echenyi$^1$ and Gyula D\'avid$^2$}

\address{$^1$ E\"otv\"os University, Department of Physics of Complex Systems,
H-1117 Budapest, P\'azm\'any P\'eter s\'et\'any 1/A , Hungary}

\address{$^2$ Department of Atomic Physics,
E{\"o}tv{\"o}s University,
H-1117 Budapest, P\'azm\'any P{\'e}ter s{\'e}t\'any 1/A, Hungary}

\begin{abstract}

The electric resistance between two arbitrary nodes on any infinite lattice structure of resistors that is a periodic tiling of space is obtained.
Our general approach is based on the lattice Green's function of the Laplacian matrix associated with the network.
We present several non-trivial examples to show how efficient our method is.
Deriving explicit resistance formulas it is shown that the Kagom\'e, the diced and the decorated lattice can be mapped
to the triangular and square lattice of resistors.
Our work can be extended to the random walk problem or to electron dynamics in condensed matter physics.

\end{abstract}


%\submitto{JPA}

\maketitle

\section{Introduction} \label{sec:intro}

In electric circuit theory one of the classic problems is to calculate the resistance between two arbitrary grid points on an
infinite square lattice of identical resistors.
This problem has been published by van der Pol and Bremmer~\cite{vanderPol}.
In a special case, it is well known that the resistance between two adjacent grid points of an infinite square lattice is
half the resistance of a bond, and an elegant and elementary solution of the problem is given by Aitchison~\cite{Aitchison}.
Over many years numerous authors have studied this problem and its extension to triangular, honeycomb and hypercube lattices
(for some relevant references see, eg,~\cite{Venezian,Atkinson,CsJ_resistor-1:cikk}).

The resistor network can systematically be treated by the Laplacian operator of
the difference equations governed by Ohm's and Kirchhoff's laws.
Then the Green's function corresponding to the discrete Laplacian operator can be related to the resistance between 
two arbitrary nodes on a resistor network.
This idea has been applied by Kirkpatrick to study percolating networks of resistors~\cite{Kirkpatrick}.
Recently, it has been demonstrated that the application of the Green's function is a very efficient way 
to find the resistance of an infinite resistor network as well~\cite{CsJ_resistor-1:cikk}.
The concept of a Green's function is widely used in the literature.
Economou's book~\cite{Economou} gives an excellent introduction to the Green's function.
Katsura~et.~al. have a review of the lattice Green's function~\cite{Katsura} and for more works on this topic,
see references in~\cite{CsJ_resistor-1:cikk}.
From the equation for the Green's function one can, in principle, derive some recurrence formulas for the resistances
between arbitrary grid points of an infinite lattice (for square lattice of resistor network see reference~\cite{CsJ_resistor-1:cikk}).
The Green's function can also be a useful tool to study, eg, capacitor networks~\cite{Asad_capacitor:cikk}
or the resistance in a perturbed lattice in which one of the bonds is missing in the lattice~\cite{CsJ_resistor-2:cikk}.
The Green's function for the anisotropic diamond lattice was discussed by Hijjawi~\cite{Hijjawi:cikk}
and the analytical properties of the Green's function in all dimensions were investigated by Guttmann~\cite{Guttman:cikk}.

The problem of a finite resistor network is equally interesting in circuit theory.
Recently, Wu has developed a theory to calculate the resistance between arbitrary nodes
for a finite lattice of resistors~\cite{Wu_finite_1:cikk},  Tzeng and Wu for impedance networks~\cite{Wu_Tzeng_finite_impedance:cikk}.
The corner-to-corner resistance and its asymptotic expansion for free boundary conditions were obtained 
by Essam and Wu~\cite{Essam_Wu_corner_corner:cikk}
and for other boundary conditions the latter was calculated by Izmailian and Huang~\cite{PhysRevE.82.011125}.

Calculation of the resistance in electric circuit theory can also be relevant to many other problems, such as
random walks~\cite{Doyle,Lovasz_random-walk:cikk} and first-passage processes~\cite{Redner:book}.
This connection is based on the fact that the difference equations for the electrical potentials
on a lattice point of a resistor network is the same as those that occur in the above mentioned problems.
Thus the resistance problem can be regarded as a problem to solve the difference equations for an infinite network.

\begin{figure}[htb]
\centering
\includegraphics[scale=0.45]{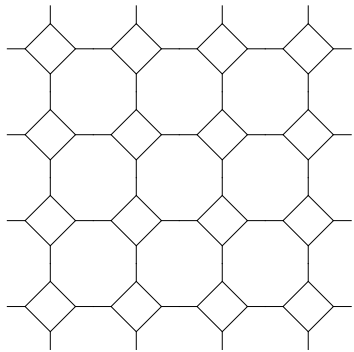}
\caption{\label{4by4_lattice:fig}
%(color online)
An example for a lattice structure of infinite resistor networks discussed in this work.
The lattice is a periodic tiling of the plane by squares and octagons.
Here, all lines represent a resistor with resistance $R$.
}
\end{figure}

In this work we generalize the Green's function approach developed in reference~\cite{CsJ_resistor-1:cikk}
for a resistor network that is a uniform tiling of $d$ dimensional space with electrical resistors.
In two dimensions, tiling is a collection of plane figures that fills the plane with no overlaps and no gaps.
Generalization to other dimensions is also possible.
Figure~\ref{4by4_lattice:fig} shows an example for a uniform tiling with resistors.
Here we derive an explicit expression for the resistance between two arbitrary nodes of any lattice structure
which is a periodic tiling of $d$ dimensional space with electrical resistors.
We present several examples of nontrivial resistor networks that, to our knowledge, have not been studied in the literature.
Our Green's function method enables us to derive, for the first time, resistance formulas for the Kagom\'e and dice lattices
(see sections~\ref{Kagome_lattice:sec} and \ref{Dice_lattice:sec}), and for the decorated lattice (see section~\ref{Lieb_lattice:sec}) in terms of the resistances on the well studied triangular and square lattices, 
respectively.
Our general method is not restricted to lattice structures made from identical resistors as is demonstrated 
in a simple, one dimensional example.
Moreover, our method can be extended to a uniform tiling of a surface of a toroid or cylinder.
The theory of perturbed lattices developed in reference~\cite{CsJ_resistor-2:cikk} can easily be extended to 
such resistor networks as we discuss in this work.

In condensed matter physics, often the tight binding model
(see, eg,~\cite{Economou,Lubensky,Ashcroft,Ziman,Kittel,Marder_Cond-Mat-Phys:book,Solyom_book_I:konyv})
is a very good approximation for calculating the electronic band structure or transport properties of crystalline matter,
like the recently discovered graphene~\cite{Novoselov_graphene-1,RevModPhys.81.109}.
In such an investigation, knowledge of the Green's function corresponding to the Schr\"odinger equation is essential.
Our results may be used to calculate the wave functions at the lattice points for complicated lattice structures.

The text is organized as follows.
In section~\ref{sec:gen_method} we outlined the general formalism to calculate the two point resistance on an arbitrary periodic lattice.
Several, non trivial examples are presented in section~\ref{sec:examples} together with a few analytical results regarding the resistance between nearby lattice points.
The conclusions are drawn in section~\ref{Conclusion:sec}.
In the Appendices we listed the resistance formulas that provide mapping of the Kagom\'e and dice lattice to the triangular lattice, and the decorated lattice to the square lattice.

\section{General formalism} \label{sec:gen_method}

Here we derive a general integral expression for the resistance between two arbitrary lattice points in an infinite regular lattice of any resitor network.

\begin{figure}[htb]
\centering
\includegraphics[scale=0.35]{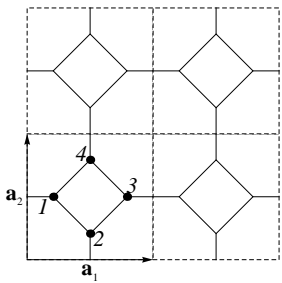}
\caption{\label{4by4_lattice_unit:fig}
%(color online)
The unit cell (dashed lines) with unit vector $\textbf{a}_1$ and $\textbf{a}_2$ of the lattice shown 
in figure~\ref{4by4_lattice:fig}.
There are four lattice points in each unit cell labeled by $\alpha = 1, 2, 3, 4$.
}
\end{figure}

First, we specify the $d$-dimensional regular lattice by its unit cell as it is common in solid state physics.
The lattice point is given by ${\bf r}= n_1 {\bf a}_1 + n_2 {\bf a}_2 + \cdots + n_d {\bf a}_d$, where
${\bf a}_1,{\bf a}_2, \dots, {\bf a}_d$ are the unit cell vectors in the $d$-dimensional space and $n_1,n_2,\dots,n_d$ are arbitrary integers.
We assume that in each unit cell there are $p$ lattice points labeled by $\alpha = 1,2, \dots, p$.
Now we denote any lattice point by $\{{\bf r},\alpha\}$, where ${\bf r}$ and $\alpha$ specify the unit cell
and the lattice point in the given unit cell, respectively.
Figure~\ref{4by4_lattice:fig} shows an example for the lattice structure of a resistor network.
We assume that all lines represent a resistor with resistance $R$ (although in the general formalism 
this restriction has not been used).
Figure \ref{4by4_lattice_unit:fig} shows one possible choice for the unit cell of the lattice shown
in figure~\ref{4by4_lattice:fig}.
There are four types of lattice point in each unit cell denoted by $\alpha = 1,2,3,4$ 
in figure~\ref{4by4_lattice_unit:fig}.

Denote the electric potential at the lattice point $\{\textbf{r},\alpha \}$ by $V_\alpha (\textbf{r})$
and the net current flowing into the network at the lattice point $\{\textbf{r},\alpha \}$  by $I_\alpha (\textbf{r})$
(as one can see below it is convenient to treat the problem by assuming that the current enter
at $\{\textbf{r},\alpha \}$ from a source outside the lattice).
Owing to Ohm's and Kirchhoff's laws the current $I_\alpha({\bf r})$ relates to the potential $V_\alpha({\bf r})$ by
the following equations:
\begin{equation}\label{I-V_law-1:eq}
    \sum_{\textbf{r}^\prime, \beta} L_{\alpha \beta}(\textbf{r}- \textbf{r}^\prime) \,  V_\beta (\textbf{r}^\prime)
    = - I_\alpha (\textbf{r}),
\end{equation}
where $L_{\alpha \beta}(\textbf{r}- \textbf{r}^\prime)$ is a $p$ by $p$ matrix called the \emph{connectivity matrix} 
or \emph{Laplacian matrix}.
For concrete examples see section~\ref{sec:examples}.
The minus sign has been introduced only for convenience.
In fact, if there is only one lattice point in each unit cell, ie, $p=1$,
then $L_{\alpha \beta}(\textbf{r}- \textbf{r}^\prime)$
is the lattice Laplacian corresponding to the finite-difference representation of
the Laplace operator used in the literature (see, eg, references~\cite{Arfken,CsJ_resistor-1:cikk}).
From the translational symmetry of the lattice structure it follows that the Laplacian matrix depends only on the difference $\textbf{r}- \textbf{r}^\prime$.
Equation~(\ref{I-V_law-1:eq}) can be written in a more compact form using matrix notation:
$\sum_{\textbf{r}^\prime} \textbf{L}(\textbf{r}- \textbf{r}^\prime) \, \textbf{V}(\textbf{r}^\prime)
    = - \textbf{I}(\textbf{r})$.
Owing to the nature of the connectivity between lattice points, the Laplacian matrix satisfies the following relation:
$\textbf{L}(\textbf{r}) = \textbf{L}^{\rm{T}} (- \textbf{r})$,
where $\rm{T}$ denotes the transpose of a matrix.

We now take periodic boundary conditions and consider a lattice with $N_1, N_2, \dots, N_d$ unit cells
along each unit cell vector $\textbf{a}_1,\textbf{a}_2, \dots, \textbf{a}_d$.
Thus the total number of unit cells in the $d$ dimensional lattice is $N = \prod_{i=1}^d N_i$ and the total number of lattice points is $Np$.
Now, it is useful to introduce the reciprocal lattice.
Owing to the periodic boundary conditions, the wave vector
${\bf k}$ in the reciprocal lattice is limited to the first Brillouin zone (BZ) and is given by
\begin{equation}
{\bf k} = \frac{m_1}{N_1}{\bf b}_1 + \frac{m_2}{N_2}{\bf b}_2
+ \cdots + \frac{m_d}{N_d} {\bf b}_d,
\label{k-vector}
\end{equation}
where $m_1,m_2,\cdots\,m_d$ are integers such that $-N_i/2 \leq m_i \leq  N_i/2$
for $i=1,2,\cdots,d$, and ${\bf b}_j$ are the reciprocal lattice vectors
defined by ${\bf a}_i {\bf b}_j = 2\pi \delta_{ij}$, $i,j=1,2,\cdots,d$.
Here we assumed that each $N_i$ is an even integer, which will be irrelevant in
the limit $N_i \rightarrow \infty$.
The mathematical description of the crystal lattice and the concept of
the Brillouin zone can be found in many books on
solid state physics~\cite{Lubensky,Ashcroft,Ziman,Kittel,Marder_Cond-Mat-Phys:book,Solyom_book_I:konyv}.

We take the discrete Fourier transform of the current $\textbf{I}(\textbf{r})$:
\numparts
\begin{eqnarray}\label{I_V_Fourier:eq}
\textbf{I}({\bf k}) &=& \sum_{{\bf r}}  \textbf{I}({\bf r})\,  \rme^{-\rmi\bf kr},
%\\
%\textbf{V}({\bf k}) &=& \sum_{{\bf r}} \textbf{V}({\bf r})\, rme^{-i\bf kr}.
\end{eqnarray}
and an analogous expression is valid for the potential $V_\alpha(\textbf{k})$.
Then the inverse Fourier transforms are given by
\begin{eqnarray} \label{I_V_inverse_Fourier:eq}
  \textbf{I}({\bf r})  &=& \frac{1}{N} \sum_{{\bf k} \in {\rm BZ}} \,
  \textbf{I}({\bf k}) \,  \rm\rme^{\rmi {\bf k}{\bf r}},
  %\\
  %\textbf{V}({\bf r})  &=& \frac{1}{N} \sum_{{\bf k} \in {\rm BZ}} \,
  %\textbf{V}({\bf k}) \,  \rm\rme^{i{\bf k}{\bf r}}.
\end{eqnarray}
and the same is valid for the potential $V_\alpha(\textbf{r})$.
\endnumparts
This expression can easily be proved using the well-known relation
$\sum_{\bf{r}} \rme^{\rmi \bf{r} \left(\bf{k} - \bf{k}^\prime \right)} = N \delta_{\bf{k},\bf{k}^\prime}$
~\cite{Lubensky,Ashcroft,Ziman,Kittel,Marder_Cond-Mat-Phys:book,Solyom_book_I:konyv}.

Using the Fourier transform of the current and the potential distributions equation~(\ref{I-V_law-1:eq}) can be rewritten as
\numparts
\begin{equation}\label{I-V_law-2:eq}
     \textbf{ L}(\textbf{k})\, \textbf{V}(\textbf{k}) = - \textbf{I}(\textbf{k}) ,
\end{equation}
where
\begin{equation}\label{Fourier_Laplace:eq}
     \textbf{L}(\textbf{k}) = \sum_\textbf{r} \textbf{L}(\textbf{r}) \, \rme^{-\rmi \bf kr}.
\end{equation}
\endnumparts
Since $\textbf{L}(\textbf{r}) = \textbf{L}^{\rm{T}} (- \textbf{r})$ it is obvious that
$\textbf{L}(\textbf{k})$ is a Hermitian matrix, ie, $\textbf{L}(\textbf{k})= \textbf{L}^+(\textbf{k}) $, where
$+$ denotes the conjugate transpose of a matrix.

For a given $\textbf{I}(\textbf{k})$ the Fourier transform $\textbf{V}(\textbf{k})$ of the potential distribution
can easily be found from (\ref{I-V_law-2:eq}):
\numparts
\begin{equation}\label{inverse_Ohm:eq}
    \textbf{V}(\textbf{k}) =  \textbf{G}(\textbf{k}) \, \textbf{I}(\textbf{k}),
\end{equation}
where the lattice \emph{Green's function} $\textbf{G}(\textbf{k})$ (in $\textbf{k}$-space) is given by
\begin{equation}\label{Green_L:eq}
    \textbf{G}(\textbf{k})= - {\textbf{L}}^{-1}(\textbf{k}).
\end{equation}
\endnumparts
Since $\textbf{L}(\textbf{k})$ is a Hermitian matrix, the Green's function $\textbf{G}(\textbf{k})$ is also a Hermitian matrix.
If there is only one lattice point in each unit cell, ie, $p=1$, the lattice Green's function $\textbf{G}(\textbf{k})$
is a one by one matrix and it is the lattice Green's function of the finite-difference representation of
the Laplace operator used in the literature (see, eg, reference~\cite{CsJ_resistor-1:cikk}).

To measure the resistance between lattice points $\{\textbf{r}_1,\alpha_1 \}$ and $\{\textbf{r}_2,\alpha_2 \}$,
we assume that a current $I$ enters at lattice point $\{\textbf{r}_1,\alpha_1 \}$ and exits at lattice point $\{\textbf{r}_2,\alpha_2 \}$,
and at any other lattice points the currents are zero.
Therefore,  the current distribution in this case can be written as
\numparts
\begin{equation}\label{init_current_dist:eq}
    I_\nu (\textbf{r}) = I \left(\delta_{\nu,\alpha_1}\, \delta_{\bf{r},\bf{r}_1} -
    \delta_{\nu,\alpha_2}\, \delta_{\bf{r},\bf{r}_2} \right),
\end{equation}
and its Fourier transform becomes
\begin{equation}\label{Fourier_init_current_dist:eq}
    I_\nu (\textbf{k}) = \sum_{\bf r} I_\nu (\textbf{r}) \, \rme^{-\rmi\bf kr}
    = I \left(\delta_{\nu,\alpha_1}\, \rme^{-\rmi \bf{k} \bf{r}_1}
    - \delta_{\nu,\alpha_2}\, \rme^{-\rmi \bf{k} \bf{r}_2}  \right).
\end{equation}
\endnumparts
Using equations~(\ref{Fourier_init_current_dist:eq}) and (\ref{inverse_Ohm:eq}) we find
\numparts
\begin{eqnarray}\label{V_k_spec:eq}
    V_\mu(\textbf{k}) &=& \sum_\nu G_{\mu \nu} I_\nu(\textbf{k})
    = I \left[ G_{\mu \alpha_1 }({\bf{k}}) \, \rme^{-\rmi {\bf{k} \bf{r}_1}}
      - G_{\mu \alpha_2}({\bf{k}}) \, \rme^{-\rmi {\bf{k} \bf{r}_2}} \right] ,
\end{eqnarray}
and the potential distribution in $\textbf{r}$-space becomes
\begin{equation}\label{V_distr-r-space:eq}
    \fl  V_\mu({\bf r}) =  \frac{1}{N} \sum_{\bf k \in \rm BZ} V_\mu({\bf k})\,  \rme^{\rmi{\bf kr}}
= \frac{I}{N}  \sum_{\bf k  \in \rm BZ}
    \left[
    G_{\mu \alpha_1 }({\bf{k}}) \, \rme^{\rmi \bf{k}\left( \bf{r} - \bf{r}_1 \right)}
      -   G_{\mu \alpha_2 }({\bf{k}}) \, \rme^{\rmi \bf{k}\left( \bf{r} - \bf{r}_2 \right)}
      \right] .
\end{equation}
\endnumparts

Now the resistance between lattice points $\{\textbf{r}_1,\alpha_1 \}$ and $\{\textbf{r}_2,\alpha_2 \}$ is given by
\begin{equation}\label{R_gen:eq}
    R_{\alpha_1 \alpha_2}(\textbf{r}_1,\textbf{r}_2) = \frac{V_{\alpha_1}(\textbf{r}_1)- V_{\alpha_2}(\textbf{r}_2)}{I}.
\end{equation}
Then using equation~(\ref{V_distr-r-space:eq}) we find
\begin{eqnarray}\label{R_final_sum_k:eq}
  R_{\alpha_1 \alpha_2}(\textbf{r}_1,\textbf{r}_2)  &=& \frac{1}{N} \, \sum_{{\bf k} \in {\rm BZ}}
    \Bigl[ G_{\alpha_1 \alpha_1}({\textbf{k}}) + G_{\alpha_2 \alpha_2}({\bf{k}}) \Bigr. \nonumber \\
&& \Bigl. -G_{\alpha_1 \alpha_2}(\textbf{k})\, \rme^{-\rmi {\bf{k}}\left( {\bf{r}}_2 - {\bf{r}}_1\right)}
    - G_{\alpha_2 \alpha_1}({\bf{k}})\, \rme^{\rmi {\bf{k}}\left( {\bf{r}}_2 - {\bf{r}}_1\right)}
    \Bigr].
\end{eqnarray}

If we take the limit $N_i \rightarrow \infty$ for all $i=1,2,\dots, d$ then the discrete summation
over ${\bf k}$ can be substituted by an integral~\cite{Lubensky,Ashcroft,Ziman,Kittel,Marder_Cond-Mat-Phys:book,Solyom_book_I:konyv}:
\begin{equation}
\frac{1}{N} \sum_{{\bf k} \in {\rm BZ}}  \rightarrow
v_0 \int_{{\bf k} \in {\rm BZ}} \, \frac{\rmd^d {\bf k}}{{(2\pi)}^d},
\label{sum-int}
\end{equation}
where $v_0$ is the volume of the unit cell.
Thus the resistance $R_{\alpha_1 \alpha_2}(\textbf{r}_1,\textbf{r}_2) $ becomes
\begin{eqnarray}\label{R_final_int_k:eq}
  R_{\alpha_1 \alpha_2}(\textbf{r}_1,\textbf{r}_2) &=&  v_0 \, \int_{{\bf k} \in {\rm BZ}} \,
    \frac{\rmd^d {\bf k}}{{(2\pi)}^d}\,
    \Bigl[ G_{\alpha_1 \alpha_1}({\bf{k}}) + G_{\alpha_2 \alpha_2}({\bf{k}}) \Bigr. \nonumber \\[2ex]
     && \Bigl.
    - G_{\alpha_1 \alpha_2}({\bf{k}})\, \rme^{-\rmi {\bf{k}}\left( {\bf{r}}_2 - {\bf{r}}_1\right)}
    - G_{\alpha_2 \alpha_1}({\bf{k}})\, \rme^{\rmi {\bf{k}}\left( {\bf{r}}_2 - {\bf{r}}_1\right)}
    \Bigr].
\end{eqnarray}
This is the central result of this work.
Note that the resistance depends only on the difference $\textbf{r}_2- \textbf{r}_1$ which is a consequence of
the translational symmetry of the resistor lattice.
Since $\textbf{G}(\textbf{k})$ is a Hermitian matrix, it is clear that the first two and the sum of the last two terms
in the above expression are real, thus the resistance is real, as it should be.
Moreover, from equation~(\ref{R_final_int_k:eq}) the following symmetry relation
$R_{\alpha_1 \alpha_2}(\textbf{r}_1, \textbf{r}_2)= R_{\alpha_2,\alpha_1}(\textbf{r}_2, \textbf{r}_1)$
is also obvious.

It is clear that deforming the lattice structure does not change the resistance between two arbitrary lattice points
if the topology of the lattice structure is preserved.
The same deformation was used by Atkinson and Steenwijk~\cite{Atkinson} for the triangular lattice.
Thus, one can always deform the lattice structure of any resistor network into a $d$ dimensional hypercube
in which all the unit cell vectors have the same magnitude and they are perpendicular to each other.
This topologically equivalent $d$ dimensional hypercube lattice is more suitable for evaluating
in equation~(\ref{R_final_int_k:eq}) the necessary integrals over the Brillouin zone
since the Brillouin zone also becomes a $d$ dimensional hypercube.
Indeed, if we write $\textbf{r}_2 - \textbf{r}_1$ in terms of the unit cell vectors:
$\textbf{r}_2 - \textbf{r}_1 = n_1 \textbf{a}_1 + n_2 \textbf{a}_2 + \cdots + n_d \textbf{a}_d$
then the exponentials in equation~(\ref{R_final_int_k:eq}) and the Green's function will depend on the variables
$x_1=\textbf{k} \textbf{a}_1, x_2=\textbf{k} \textbf{a}_2, \dots, x_d=\textbf{k} \textbf{a}_d $, ie,
\begin{equation}\label{G_k_replace_x;eq}
    \textbf{G}(x_1,\dots,x_d) =
\textbf{ G} (\textbf{k} \textbf{a}_1 \rightarrow x_1,\dots,\textbf{k} \textbf{a}_d \rightarrow x_d).
\end{equation}
Moreover, the integration variables $\textbf{k}=(k_1,k_2,\dots,k_d)$ in (\ref{R_final_int_k:eq})
can also be transformed to the variables $x_1, x_2, \dots, x_d$ for which the limits of the integration are $\pm \pi$.
The Jacobian corresponding to the transformations of variables cancels the volume $v_0$ of the unit cell
in (\ref{R_final_int_k:eq}).
Thus, the general result (\ref{R_final_int_k:eq}) for the resistance can be rewritten as a $d$ dimensional integral
that is a more suitable form for explicit calculations:
\begin{eqnarray}\label{Rabm1md_int_x1_xd:eq}
  \fl  R_{\alpha_1 \alpha_2}(n_1,\dots,n_d) =
    \int_{-\pi}^{\pi} \frac{\rmd x_1}{2\pi} \cdots
    \int_{-\pi}^{\pi} \frac{\rmd x_d}{2\pi}\, \Bigl[  G_{\alpha_1 \alpha_1}(x_1,\dots,x_d)
         + G_{\alpha_2 \alpha_2}(x_1,\dots,x_d) \Bigr.   \nonumber \\[1ex]
\fl     \Bigl.
    - G_{\alpha_1 \alpha_2}(x_1,\dots,x_d)\, \rme^{-\rmi \left(n_1 x_1+n_2 x_2 +\cdots n_d x_d \right)}
- G_{\alpha_2 \alpha_1}(x_1,\dots,x_d)\, \rme^{\rmi \left(n_1 x_1+n_2 x_2 +\cdots n_d x_d \right)} \Bigr] .
\end{eqnarray}
For any infinite regular resistor network the above result is an integral representation of the resistance between arbitrary lattice points
and provides a practical way for determining the resistance in question.

\section{Applications of the general formalism} \label{sec:examples}
In this section we demonstrate how this result can be applied to different lattice structures.
We present examples of one, two and three dimensional cases as well.

\subsection{One dimensional resistor structure}\label{1dim_r1_r2:sec}

\begin{figure}[htb]
\centering
\includegraphics[scale=0.8]{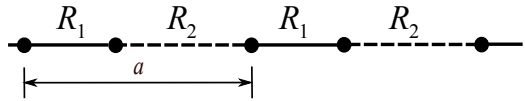}
\caption{\label{1_dim_lattice:fig}
%(color online)
A periodic tiling in one dimension with two resistors $R_1$ and $R_2$.
In this lattice each unit cell contains two non-equivalent lattice points labeled by $\alpha = 1,2$, ie, $p=2$.
The length of the unit cell is $a$.
}
\end{figure}

Perhaps, the simplest periodic tiling is the one dimensional lattice structure in which each unit cell contains two non-equivalent resistors
as shown in figure~\ref{1_dim_lattice:fig}.
This lattice structure is also an example of the case in which the connectivity between lattice points in each unit cell
is not necessary the same.
As we mentioned before, our general formalism outlined above can be applied to such cases as well.
Indeed, now there are two different resistances in the unit cell, $R_1$ and $R_2$.

In one dimension we can characterize lattice points by $\{r,\alpha\}$, where $r = m a$ and $\alpha =1,2$
(here $m$ is an integer).
Ohm's and Kirchhoff's laws for the currents $I_\alpha(r)$
at site $\{r,\alpha\}$ (with $\alpha =1,2$) can be written as
\numparts
\begin{eqnarray}
 I_1(r) & = &  \frac{ V_1(r)-V_2(r)}{R_1}   +
 \frac{ V_1(r)-V_2(r-a)}{R_2} ,   \\[1ex]
I_2(r) & = & \frac{ V_2(r)-V_1(r)}{R_1}   +
  \frac{ V_2(r)-V_1(r+a)}{R_2}   .
\label{aram-1d_r_repi:eq}
\end{eqnarray}
\endnumparts
Hence, the Laplacian matrix can be written as
\numparts
\begin{eqnarray}
\label{1d_Lm-r-space:eq}
\mathbf{L}(r) &=&
\left[\begin{array}{cc}
 -\frac{\delta_{r,0}}{R_1}-\frac{\delta_{r,0}}{R_2} & \frac{\delta_{r,0}}{R_1}+\frac{\delta_{r,a}}{R_2} \\[2ex]
 \frac{\delta_{r,0}}{R_1}+\frac{\delta_{r,-a}}{R_2} & -\frac{\delta_{r,0}}{R_1}-\frac{\delta_{r,0}}{R_2}
\end{array}\right],
\end{eqnarray}
and its Fourier transformation reads
\begin{eqnarray}
\label{1d_Lm:eq}
\mathbf{L}(k) &=&
\left[\begin{array}{cc}
-\frac{1}{R_1}-\frac{1}{R_2}& \frac{1}{R_1}+\frac{\rme^{-\rmi k a}}{R_2}\\[2ex]
\frac{1}{R_1}+\frac{\rme^{\rmi k a}}{R_2} & -\frac{1}{R_1}-\frac{1}{R_2}
\end{array}\right].
\end{eqnarray}
\endnumparts
Inverting the matrix $\textbf{L}(k)$ the Green's function defined by equation~(\ref{Green_L:eq}) becomes
\begin{eqnarray}
\label{1d_Gm:eq}
\hspace{-4mm} \mathbf{G}(k) &=&
\frac{1}{4 \sin^2 \frac{k a}{2} }
\left[
\begin{array}{cc}
   R_1+R_2  & R_2+ R_1 \rme^{-\rmi k a}  \\[2ex]
 R_2+ R_1 \rme^{-\rmi k a}  & R_1+R_2
\end{array}
\right]\! .
\end{eqnarray}
Now, changing the variable $k$ to $x$, the Green's function in equation~(\ref{Rabm1md_int_x1_xd:eq}) (for $d=1$)
transforms to  $G(x)=G(ka \to x)$.

The resistance between lattice points $\{0,\alpha\}$ and $\{m a,\beta\}$ can be obtained
from equation~(\ref{Rabm1md_int_x1_xd:eq}) for the one-dimensional case ( $d=1$):
\numparts
\begin{eqnarray}\label{R11m_:eq}
R_{11}(m) &=& R_{22}(m) = \left(R_1 +R_2 \right) f(m) , \\[2ex]
R_{12}(m) &=&  R_1 f(m+1) + R_2 f(m) ,\\[2ex]
R_{21}(m) &=&  R_1 f(m-1) + R_2 f(m) ,   \,\,\, \rm{where} \\[2ex]
f(m) &=& \int_{-\pi}^{\pi} \frac{d x}{2\pi} \frac{ 1 - \cos m x}{1 - \cos x}.
\end{eqnarray}
\endnumparts
The function $f(m)$ can be evaluated using the method of residues and we find $f(m) = |m|$, if $m$ is an integer.
Thus, $R_{11}(m) = \left( R_1 +R_2\right) |m|$, $R_{12}(m) =  R_1 |m+1| + R_2 |m| $ and
$R_{21}(m) =  R_1 |m-1| + R_2 |m|$.
These results can easily be understood from the fact that the current flows only between lattice point
$\{0,\alpha\}$ and $\{m a,\beta\}$, and between these points the resistances are in series.
The two semi-infinite segments of the chain do not affect the resistance.

\subsection{Tiling of the plane by squares and octagons}\label{sq_oct:sec}

Consider the resistor network shown in figure~\ref{4by4_lattice:fig}.
Hereafter, we assume that all lines represent a resistor with resistance $R$.
Using Ohm's and Kirchhoff's laws the currents $I_\alpha(\textbf{r})$
at site $\{\textbf{r},\alpha\}$ (with $\alpha = 1,2,3,4$) can be written as
\numparts
\begin{eqnarray}
 I_1({\bf r}) & = &  \frac{ V_1(\textbf{r})-V_2(\textbf{r})}{R}   +
 \frac{ V_1(\textbf{r})-V_3(\textbf{r}-\textbf{a}_1)}{R}  + \frac{ V_1(\textbf{r})-V_4(\textbf{r})}{R}, \\[2ex]
I_2({\bf r}) & = & \frac{ V_2(\textbf{r})-V_1(\textbf{r})}{R}   +
  \frac{ V_2(\textbf{r})-V_3(\textbf{r})}{R} + \frac{ V_2(\textbf{r})-V_4(\textbf{r}-\textbf{a}_2)}{R}, \\[2ex]
I_3({\bf r}) & = & \frac{ V_3(\textbf{r})-V_1(\textbf{r}+\textbf{a}_1)}{R}   +
 \frac{ V_3(\textbf{r})-V_2(\textbf{r})}{R}  + \frac{ V_3(\textbf{r})-V_4(\textbf{r})}{R},  \\[2ex]
I_4({\bf r}) & = & \frac{ V_4(\textbf{r})-V_1(\textbf{r})}{R}   +
 \frac{ V_4(\textbf{r})-V_2(\textbf{r}+\textbf{a}_2)}{R}  + \frac{ V_4(\textbf{r})-V_3(\textbf{r})}{R}.
\label{aram-square_r_repi:eq}
\end{eqnarray}
\endnumparts
From these equations one can easily read out the Laplacian matrix:
\begin{equation}\label{L_4by4_r_space:eq}
  \textbf{L}({\bf r}) =  \frac{1}{R} \,  \left [ \begin{array} {cccc}
-3 \delta_{{\bf r},{\bf r}}  & \delta_{{\bf r},{\bf 0}} & \delta_{{\bf r},{\bf a}_1} & \delta_{{\bf r},{\bf 0}} \\[1ex]
\delta_{{\bf r},{\bf 0}} & -3 \delta_{{\bf r},{\bf 0}}  & \delta_{{\bf r},{\bf 0}}  & \delta_{{\bf r},{\bf a}_2} \\[1ex]
\delta_{{\bf r},-{\bf a}_1}  & \delta_{{\bf r},{\bf 0}}  & -3 \delta_{{\bf r},{\bf 0}}  & \delta_{{\bf r},{\bf 0}}  \\[1ex]
\delta_{{\bf r},{\bf 0}} & \delta_{{\bf r},-{\bf a}_2} & \delta_{{\bf r},{\bf 0}}  & -3 \delta_{{\bf r},{\bf 0}}
\end{array}  \right ].
\end{equation}
Hence, the Fourier transform of the Laplacian matrix (\ref{L_4by4_r_space:eq}) is given by
\begin{equation}\label{L_4by4:eq}
  \textbf{L}(\textbf{k}) = \frac{1}{R} \,  \left [ \begin{array} {cccc}
-3  & 1 & \rme^{-\rmi {\bf k} {\bf a}_1 } & 1 \\[1ex]
1  & -3 & 1 & \rme^{-\rmi {\bf k} {\bf a}_2 } \\[1ex]
\rme^{\rmi {\bf k} {\bf a}_1 }  & 1 & -3 & 1 \\[1ex]
1  & \rme^{\rmi {\bf k} {\bf a}_2 } & 1 & -3
\end{array}  \right ].
\end{equation}

Inverting the above matrix  $\textbf{L}(\textbf{k})$,  the Green's function can easily be calculated from equation~(\ref{Green_L:eq})
and changing the variables we find
\begin{equation}\label{G_replace_2dim:eq}
    \textbf{G}(x_1,x_2) =\textbf{ G} (\textbf{k} \textbf{a}_1 \rightarrow x_1,\textbf{k} \textbf{a}_2 \rightarrow x_2).
\end{equation}
Since the resistance depends only on the difference  $\textbf{r}_2 - \textbf{r}_1$
we can take the lattice vector $\textbf{r}_1$ (at which the current $I$ enters the network) to the origin.
Specify the lattice vector $\textbf{r}_2 \equiv \textbf{r}_0 = m \textbf{a}_1 + n \textbf{a}_2 $
(at which the current $I$ exits the network).
Then using the main result (\ref{Rabm1md_int_x1_xd:eq}) for $d=2$,
the resistance between lattice point $\{\textbf{0},\alpha\}$ and $\{\textbf{r}_0,\beta\}$ is given by
\begin{eqnarray}\label{Rabmn_int_x1_x2:eq}
       R_{\alpha \beta}(m,n) &=&
    \int_{-\pi}^{\pi} \frac{\rmd x_1}{2\pi}\,  \int_{-\pi}^{\pi} \frac{\rmd x_2}{2\pi}
    \Bigl[ G_{\alpha \alpha}(x_1,x_2) + G_{\beta \beta}(x_1,x_2)  \Bigr.  \nonumber \\[2ex]
    && - G_{\alpha \beta}(x_1,x_2) \rme^{-\rmi \left(m x_1+n x_2 \right)}
       - G_{\beta \alpha}(x_1,x_2) \rme^{\rmi \left(m x_1+n x_2 \right)} \Bigr].
\end{eqnarray}
It can be shown that in the above expression one integral can be evaluated analytically by the method of residues~\cite{Arfken}
much in the same way as in reference~\cite{CsJ_resistor-1:cikk} for the square and triangular lattices.

Applying this result we calculated the resistance for a few cases.
For example, for the resistance between lattice point $1$ and $2$ that belong to the same unit cell, we obtain
\numparts
%\begin{widetext}
\begin{eqnarray}\label{r1200_4by4:eq}
  \hspace{-7mm}  R_{12}(0,0) &=&  \frac{R}{2}\, \int_{-\pi}^{\pi} \frac{\rmd x_1}{2\pi}
\int_{-\pi}^{\pi} \frac{\rmd x_2}{2\pi}\, f(x_1,x_2),  \nonumber \\
\hspace{-7mm}  f(x_1,x_2) \! &=& \! \frac{-9 + 4 \cos x_1 + 4 \cos x_2 + \cos \left(x_1 - x_2\right) }
 {-7 + 3 \cos x_1  + 3 \cos x_2  + \cos x_1 \cos x_2 } \! .
 \end{eqnarray}
Performing the integrations we find
\begin{eqnarray}\label{r1200_4by4_res:eq}
 R_{12}(0,0) &=& \left(\frac{1}{2} + \frac{\sqrt{2} \arctan \left(2\sqrt{2}\right)}{4\pi}\right) R,
\end{eqnarray}
and numerically, $ R_{12}(0,0)  \approx 0.6385 R$.

Similarly, the resistance between lattice point $1$ and $3$ (when they are in the same unit cell) is:
\begin{eqnarray}\label{r1300_4by4:eq}
  \hspace{-8mm}  R_{13}(0,0) &=& \! \frac{3 \sqrt{2}}{2} \left( 1- \frac{2 \arctan \sqrt{2}}{\pi}\right) R \approx 0.8312 R.
\end{eqnarray}
%\end{widetext}
Owing to the symmetry of the lattice, $R_{14}(0,0) = R_{23}(0,0) = R_{34}(0,0)=R_{12}(0,0)$, and
$R_{24}(0,0) = R_{13}(0,0)$.

In adjacent unit cells the two squares are connected by one resistor, and then one can ask
what is the resistance between the two ends of this resistor.
In our notation, this can be found by calculating, eg, $R_{31}(1,0)$.
Using  equation~(\ref{Rabmn_int_x1_x2:eq}) we obtain
\begin{eqnarray}\label{r3110_4by4:eq}
  R_{31}(1,0) &=& \left(1 -\frac{\sqrt{2}}{2} + \frac{\sqrt{2}\arctan \sqrt{2}}{\pi}\right) R,
\end{eqnarray}
%\end{widetext}
\endnumparts
and numerically, $ R_{31}(1,0)  \approx 0.7229 R$.
For more examples, it is useful to apply the method of residues.

%(-5 + Cos[x2] + Cos[x1] (3 + Cos[x2]))/(-7 + 3 Cos[x2] + Cos[x1] (3 + Cos[x2]))
%1 - 1/Sqrt[2] + (Sqrt[2] ArcTan[Sqrt[2]])/\[Pi] %0.722937

\subsection{The Kagom\'e lattice}\label{Kagome_lattice:sec}
\begin{figure}[htb]
\centering
\includegraphics[scale=0.6]{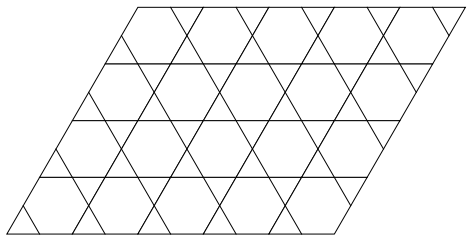}
\caption{\label{Kagome_lattice:fig}
%(color online)
The Kagom\'e lattice structure of the resistor network.
}
\end{figure}

\begin{figure}[htb]
\centering
\includegraphics[scale=0.4]{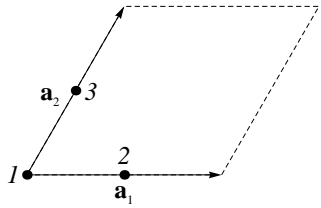}
\caption{\label{Kagome_unitcell_lattice:fig}
%(color online)
The unit cell of the Kagom\'e lattice with three lattice points, $p=3$.
}
\end{figure}

\begin{figure}[htb]
\centering
\includegraphics[scale=0.4]{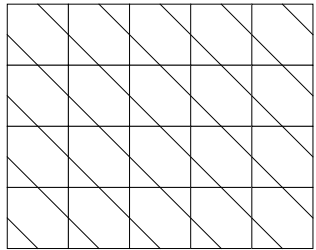}
\caption{\label{Kagome_transformed_lattice:fig}
%(color online)
The topologically equivalent lattice structure of the Kagom\'e lattice.
The unit cell is a square.
}
\end{figure}

The next example we consider is the so-called Kagom\'e lattice structure~\cite{Solyom_book_I:konyv} shown in figure~\ref{Kagome_lattice:fig}.
The unit cell of the Kagom\'e lattice structure is shown in figure~\ref{Kagome_unitcell_lattice:fig}.
Each unit cell contains three lattice points, ie, $\alpha = 1,2,3$ and
the Laplacian matrix is a three by three matrix.
In a similar way as in section~\ref{sq_oct:sec}, one can write down the equations for the currents 
for each lattice point in one unit cell.
Then it is easy to show that the Fourier transform of the Laplacian matrix for the Kagom\'e lattice is
\begin{equation}\label{L_Kagome:eq}
  \textbf{L}(\textbf{k}) = \frac{1}{R} \, \left [ \begin{array} {ccc}
-4  & 1+ \rme^{-\rmi {\bf k} {\bf a}_1} & 1+ \rme^{-\rmi {\bf k} {\bf a}_2}  \\[1ex]
1+ \rme^{\rmi {\bf k} {\bf a}_1 }  & -4 & 1+ \rme^{\rmi {\bf k}\left( {\bf a}_1 - {\bf a}_2 \right)} \\[1ex]
1+ \rme^{\rmi {\bf k} {\bf a}_2 }  & 1+ \rme^{-\rmi {\bf k}\left( {\bf a}_1 - {\bf a}_2 \right) }& -4
\end{array}  \right ].
\end{equation}
As we mentioned before equation~(\ref{G_k_replace_x;eq}) the lattice can be deformed to a topologically equivalent one
in which the unit cell is a hypercube, in this case, it is a square lattice (see figure~\ref{Kagome_transformed_lattice:fig}).
Now, calculating the Green's function from equation~(\ref{Green_L:eq}) and changing variables according to equation~(\ref{G_replace_2dim:eq})
one can use the general expression (\ref{Rabmn_int_x1_x2:eq}) to find the resistance between arbitrary lattice points 
on the Kagom\'e resistor network.

It is interesting to calculate the resistance between adjacent lattice points.
From symmetry it follows that $R_{12}(0,0)=R_{13}(0,0)=R_{23}(0,0)$, and from equation~(\ref{Rabmn_int_x1_x2:eq}) we find that
the integrand of the sum of these resistances is a constant and equals $3R/2$.
Therefore, in equation~(\ref{Rabmn_int_x1_x2:eq}) the double integral results in $R_{12}(0,0)+R_{13}(0,0)+R_{23}(0,0) = 3R/2$ and
the resistance between the nearest neighbor lattice points is $R_{12}(0,0)=R_{13}(0,0)=R_{23}(0,0)= R/2$.
Based on the superposition principle and the symmetry of the lattice structure the same result was found
in references~\cite{Bartis_Kagome:cikk,Osterberga_Kagome:cikk}.

However, our general formalism enables us to find the resistance between arbitrary lattice points.
Moreover, we show that the resistance $R_{\alpha \beta}(m,n)$ on the Kagom\'e lattice can be expressed
in terms of the resistances $R^\vartriangle(m,n)$ on a triangular lattice.
To this end, we recall the result for the triangular lattice as discussed in references~\cite{Atkinson,CsJ_resistor-1:cikk}.
The resistance  between lattice points $\{\textbf{0},1\}$ and $\{\textbf{r}_0,1\}$
(here $\textbf{r}_0 = m \textbf{a}_1 + n \textbf{a}_2$ and for the triangular lattice there is only one lattice point
in each unit cell which is labeled here by $1$) is given by
\begin{eqnarray}\label{R_3szog:eq}
R^\vartriangle(m,n)&=& R \int_{-\pi}^{\pi} \frac{\rmd x_1}{2\pi}
\int_{-\pi}^{\pi} \frac{\rmd x_2}{2\pi} f_{mn}^\vartriangle(x_1,x_2), \,\,\, \textrm{where }\nonumber \\[2ex]
f_{mn}^\vartriangle(x_1,x_2) &=& \frac{1- \cos(m x_1+ n x_2)}{3-\cos x_1 - \cos x_2 - \cos(x_1 - x_2)}.
\end{eqnarray}
The resistor problem for triangular lattice has been well studied in references~\cite{Atkinson,CsJ_resistor-1:cikk}
including a few analytical results.
Note that the above expression is slightly different from that in reference~\cite{CsJ_resistor-1:cikk}
since here the angle between the two unit cell vectors $\textbf{a}_1$ and $\textbf{a}_2$ is $60^\circ$,
while in reference~\cite{CsJ_resistor-1:cikk} it is  $120^\circ$, but this different choice of the unit cell is irrelevant 
for the resistances.
Now, starting from equation~(\ref{Rabmn_int_x1_x2:eq}) one can show that the resistance $R_{\alpha \beta}(m,n)$
on the Kagom\'e lattice can be expressed in terms of the resistances $R^\vartriangle(m,n)$ on the triangular lattice.
The explicit expressions are given in~\ref{kagome_3szog:app}.

Such an interesting map between the Kagom\'e and the triangular resistor networks can be understood qualitatively by applying  
the so-called triangular-star transformation to the Kagom\'e  lattice just as it was used by Atkinson and Steenwijk 
for honeycomb lattices~\cite{Atkinson}. 
However, the Green's function method provides a more systematic way to deduce the resistance formulas as it was shown 
for the honeycomb lattice in reference~\cite{CsJ_resistor-1:cikk}. 
Indeed, mathematically the map between the Kagom\'e and the triangular resistor networks is based on the fact that
the determinant of the Laplacian (\ref{L_Kagome:eq}) for the Kagom\'e lattice differs from the denominator of the function $f_{mn}^\vartriangle(x_1,x_2)$ 
only by a constant factor (as it was the case for honeycomb lattice~\cite{CsJ_resistor-1:cikk}).
This mathematical fact should have a topological origin, which can be a research topic in the future.

Using these results (see~\ref{kagome_3szog:app}) it is easy to find the resistance
between the opposite corners of a hexagon on the Kagom\'e lattice, eg, by calculating $R_{33}(1,0)$ or $R_{22}(0,1)$
and we obtain $R_{33}(1,0) = R_{22}(0,1) = \left(\frac{4}{9}+\frac{2\sqrt{3}}{3\pi} \right) R \approx 0.8120 R$.
% 0.811997

\subsection{The dice lattice}\label{Dice_lattice:sec}

\begin{figure}[htb]
\centering
\includegraphics[scale=0.65]{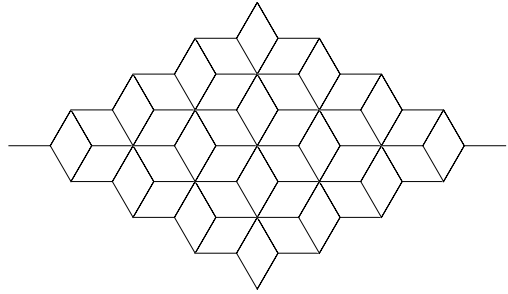}
\caption{\label{Dice_lattice:fig}
%(color online)
The dice lattice structure.
}
\end{figure}

\begin{figure}[htb]
\centering
\includegraphics[scale=0.35]{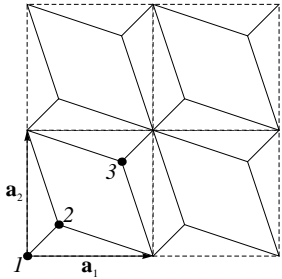}
\caption{\label{Dice_trafo_lattice:fig}
%(color online)
The lattice structure that is topologically equivalent to the dice lattice structure.
}
\end{figure}

The dice lattice (figure~\ref{Dice_lattice:fig}) is a periodic tiling of the
plane by rhombi having $60^\circ$ and $120^\circ$ interior angles and all vertices
have degree $3$ or $6$.
The dice lattice is a frequently used lattice structure in the literature.
Recently antiferromagnetic behavior has been studied on the dice lattice~\cite{PRL_101_030601_2008:cikk}.
%The dice lattice is the dual of the Kagom\'e lattice, which is in turn the medial graph of the triangular and hexagonal lattices.
\begin{figure}[htb]
\centering
\includegraphics[scale=0.5]{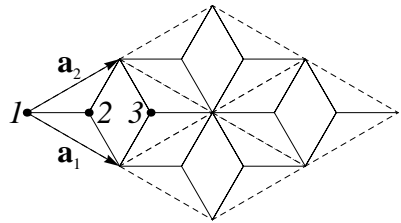}
\caption{\label{Dice_unitcell:fig}
%(color online)
The unit cell of the dice lattice with three lattice points, $p=3$.
}
\end{figure}
One possible choice of the unit cell is shown in figure~\ref{Dice_unitcell:fig}, it contains three lattice points.
The dice lattice can be deformed to a topologically equivalent, two dimensional square lattice
as shown in figure~\ref{Dice_trafo_lattice:fig}.

Using the method outlined previously one can easily obtain the Fourier transform of the Laplacian matrix
for dice lattice and it is given by
\begin{equation}\label{L_Dice:eq}
  \textbf{L}(\textbf{k}) = \frac{1}{R} \, \left [ \begin{array} {ccc}
-6  & A^* & B^* \\[1ex]
A & -3 & 0 \\[1ex]
B & 0 & -3
\end{array}  \right ],
\end{equation}
where $A =  1+ \rme^{\rmi {\bf k} {\bf a}_1} + \rme^{\rmi {\bf k} {\bf a}_2}$ and
$B = \rme^{\rmi {\bf k} {\bf a}_1} + \rme^{\rmi {\bf k} {\bf a}_2} + \rme^{\rmi {\bf k}\left( {\bf a}_1 + {\bf a}_2 \right)}$.

Now, one can again use equation~(\ref{Rabmn_int_x1_x2:eq}) to calculate the resistance between arbitrary lattice points 
on a dice lattice.
Much in the same way as in the case of Kagom\'e lattice from equation~(\ref{Rabmn_int_x1_x2:eq}) we find that
the resistance between arbitrary lattice points on a dice lattice can again be expressed
in terms of the corresponding resistance on the triangular lattice.
The explicit expressions are given in~\ref{diced_3szog:app}.
The mathematical reason for this mapping is the same as in the case of Kagom\'e lattice mentioned in section~\ref{Kagome_lattice:sec}.

Here we present a few analytical results.
For instance,  the resistance between lattice point $1$ and $2$ (which are in the same unit cell) may be found by calculating $R_{12}(0,0)$ and we obtain
a simple result $R_{12}(0,0) = R/2 $.
Moreover, the resistance between the two ends of the shorter and longer diagonal of a rhombus on a dice lattice
may be obtained by calculating $R_{23}(0,0)$ and $R_{11}(1,0)$, respectively and
we find $R_{23}(0,0) = \left(\frac{5}{9}+ \frac{\sqrt{3}}{3\pi}\right)R \approx 0.7393 R$ and $R_{11}(1,0)= R/2$.

\subsection{The decorated lattice}\label{Lieb_lattice:sec}
\begin{figure}[htb]
\centering
\includegraphics[scale=0.4]{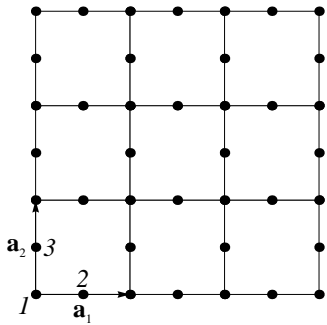}
\caption{\label{Lieb_lattice:fig}
%(color online)
The decorated lattice structure.
Between each pair of full circles the bond represents a resistor with resistance $R$.
The unit cell contains three lattice points labeled by $\alpha=1,2,3$, ie, $p=3$.
}
\end{figure}

Here we consider the so-called decorated which is a square lattice with extra lattice points in the middle of each bond as shown in figure~\ref{Lieb_lattice:fig}
together with a natural choice for the unit cell vectors and lattice points labeled by $\alpha=1,2,3$ in each unit cell.
This lattice was used to study, eg, the phase diagram of a decorated Ising system~\cite{Wu_PhysRevB.8.4219}
and recently the Hubbard model~\cite{Lieb_PRL_62_1201_1989:cikk}.

In much the same way as in our earlier examples one can easily find
the Fourier transform of the Laplacian matrix for the decorated lattice:
\begin{equation}\label{L_Lieb:eq}
  \textbf{L}(\textbf{k}) = \frac{1}{R} \, \left [ \begin{array} {ccc}
-4  & 1+ \rme^{-\rmi {\bf k} {\bf a}_1 } & 1+ \rme^{-\rmi {\bf k} {\bf a}_2 }  \\[1ex]
1+ \rme^{\rmi {\bf k} {\bf a}_1 }  & -2 & 0 \\[1ex]
1+ \rme^{\rmi {\bf k} {\bf a}_2 }  & 0 & -2
\end{array}  \right ].
\end{equation}
Then, the resistance between arbitrary lattice points can be obtained from equation~(\ref{Rabmn_int_x1_x2:eq}).
However, again just as in the case of the Kagom\'e and dice lattices here there is a close relation between
the decorated lattice and the square lattice regarding the resistances.
We derived explicit expressions (see~\ref{Lieb_4szog:app}) for the resistance $R_{\alpha \beta}(m,n)$ on a decorated lattice
in terms of the resistances $R^\square(m,n)$ (between the origin and the lattice point $ m \textbf{a}_1 + n \textbf{a}_2$ on a square lattice) given by
\begin{eqnarray}\label{R_4szog:eq}
R^\square(m,n)&=& R \int_{-\pi}^{\pi} \frac{\rmd x_1}{2\pi}
\int_{-\pi}^{\pi} \frac{\rmd x_2}{2\pi} f_{mn}^\square(x_1,x_2), \,\,\, \textrm{where }\nonumber \\[2ex]
f_{mn}^\square(x_1,x_2) &=& \frac{1- \cos(m x_1+ n x_2)}{2-\cos x_1 - \cos x_2}.
\end{eqnarray}
Regarding the resistor problem on square lattice see, eg, references~\cite{Atkinson,CsJ_resistor-1:cikk} in which a few analytical results~\cite{Atkinson}
and recurrence formulas~\cite{CsJ_resistor-1:cikk} are presented.
The mathematical reason for the mapping of the decorated lattice to the square lattice of resistor network is again 
based on the fact that the determinant $\det({\bf L})$ of the Laplacian (\ref{L_Lieb:eq}) differs 
from the denominator of the function $f_{mn}^\square(x_1,x_2)$ only by a constant factor, 
similarly as in the case of Kagom\'e lattice mentioned in section~\ref{Kagome_lattice:sec}. 

Note that the first equation in~\ref{Lieb_4szog:app} can easily be understood by a simple argument.
To find the resistance $R_{11}(m,n)$ one can disregard the middle points of each bond on the decorated lattice.
Then the decorated lattice is equivalent to a square lattice in which the bonds have resistance  $2R$.
Thus we find that $R_{11}(m,n)=2R^\square(m,n) $ in agreement with that given by the first equation in~\ref{Lieb_4szog:app}.
For example, $R_{11}(1,1)= 4R/\pi$.

Using the results given in~\ref{Lieb_4szog:app}, we find that the resistance
between lattice points $1$ and $2$ that belong to the same unit cell is $R_{12}(0,0)=3R/4$.
Similarly, the resistance between lattice points $2$ and $3$
(which are in the same unit cell) is $R_{23}(0,0) = \left(1+ 1/\pi \right)R \approx 1.3183 R$.

\subsection{Centered square lattice}\label{centered_sq_lattice:sec}

\begin{figure}[htb]
\centering
\includegraphics[scale=0.35]{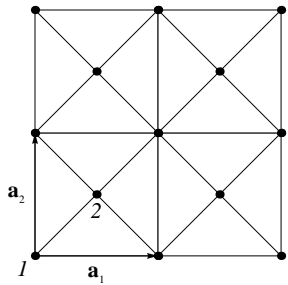}
\caption{\label{cent_sq_lattice:fig}
%(color online)
The centered square lattice structure.
Between each full circles the bond represents a resistor with resistance $R$.
The unit cell contains two lattice points labeled by $\alpha=1,2$, ie, $p=2$.
}
\end{figure}

In sections~\ref{Kagome_lattice:sec} and \ref{Dice_lattice:sec}
we showed that the Kagom\'e and the dice lattice can be mapped to the triangular lattice,
while in section~\ref{Lieb_lattice:sec} it was shown that the decorated lattice is mapped to the square lattice.
In this section we show another interesting mapping.
In particularly, we find that the resistor network discussed in section~\ref{sq_oct:sec} (see figure~\ref{4by4_lattice:fig})
can be mapped to the so-called centered square lattice shown in figure~\ref{cent_sq_lattice:fig}.

Using Ohm's and Kirchhoff's laws we find easily that the Fourier transform of the Laplacian is
\numparts
\begin{eqnarray}
\label{c_sq_Lm:eq}
\mathbf{L}(\mathbf{k}) &=&
\frac{1}{R} \left[ \begin{array}{cc}
A & B^*\\
B & -4
\end{array}
\right], \rm{where} \\[1ex]
A &=&-8 + 2\left(\cos {\bf k} {\bf a}_1 + \cos {\bf k} {\bf a}_2\right),  \\[1ex]
B &=&\left(1+\rme^{\rmi {\bf k} {\bf a}_1} \right)\left(1+\rme^{\rmi {\bf k} {\bf a}_2}\right).
\end{eqnarray}
\endnumparts
Then, the Green's function from equation~(\ref{Green_L:eq}) becomes
\begin{equation}
\mathbf{G}(\mathbf{k})=\frac{1}{\det (\textbf{L})}
\left[\begin{array}{cc}
4 & B^*\\
B & -A
\end{array}\right],
\end{equation}
where the determinant of $\textbf{L}(\textbf{k})$ is given by
\begin{eqnarray}
\label{c_sq_detL:eq}
R \det (\mathbf{L}) &=& 28-12( \cos {\bf k} {\bf a}_1 + \cos {\bf k} {\bf a}_2 )
\nonumber \\[1ex]
&& - 4 \cos {\bf k} {\bf a}_1  \cos {\bf k} {\bf a}_2.
\label{eq:c_sq determinans}
\end{eqnarray}
% 28 - 12 Cos[x1] - 12 Cos[x2] - 4 Cos[x1] Cos[x2]
Note that this determinant is exactly the same as that of the Laplacian in equation~(\ref{L_4by4:eq}).
Therefore, the Green's function should have the same denominator in both cases.
The numerator of the Green's functions in the two cases is different.
However, the structure of the numerator of the Green's function for $\textbf{L}(\textbf{k})$ given in (\ref{L_4by4:eq}) 
is the same as that for the centered square lattice.
Thus, one can work out the resistance formulas (not presented here) analogous to those listed in the Appendices
to find the explicit mapping between the lattice structure shown in figures~\ref{4by4_lattice:fig} and \ref{cent_sq_lattice:fig}.

Using equation~(\ref{Rabmn_int_x1_x2:eq}) we calculated the resistance analytically for a few cases:
\numparts
\begin{eqnarray}\label{r_csq_res:eq}
 \hspace{-7mm} R_{11}(1,0) &=& \frac{\sqrt{2}  \arctan \left(\sqrt{2}/2\right)}{\pi} R, \\[2ex]
 \hspace{-7mm} R_{12}(0,0) &=& \Bigl[\frac{1}{2} - \frac{\sqrt{2}}{4\pi}\, \arctan \left(2\sqrt{2}\right)\Bigr] R, \\[2ex]
 R_{22}(1,0)  &=& \Bigl[ -1 + \frac{1}{\pi} +  \frac{9\sqrt{2}}{4\pi}\, \arctan \left(2\sqrt{2}\right)\Bigr] R,
\end{eqnarray}
\endnumparts
and numerically, $ R_{11}(1,0) \approx 0.2771 R$, $ R_{12}(0,0) \approx 0.3615 R$ and $R_{22}(1,0) \approx 0.5651 R$.
% R_{11}(1,0) = Sqrt[2] ArcTan[1/Sqrt[2]])/\[Pi], 0.277063
% R_{12}(0,0) = 1/4 (2 - Sqrt[2] + (2 Sqrt[2] ArcTan[Sqrt[2]])/\[Pi]), 0.361468
% egyszerubb R_{12}(0,0) = 1/2 - Sqrt[2]/(4\pi) ArcTan[2 Sqrt[2]]] \approx 0.361468
% R_22(1,0) = -1 + 1/\[Pi] + (9 Sqrt[2] ArcTan[Sqrt[2]/2])/(2 \[Pi]), 0.565094
%% AZONOSSAG  ArcTan (a) + ArcTan(b) = ArcTan((a+b)/(1- ab))
%%   2 ArcTan (x) = ArCtan ((2x)/(1-x^2))
%%  2 ArcTan (\sqrt{2}/2) = ArcTan (2 \sqrtt{2})

\subsection{Tiling of the plane by squares and triangles}\label{snub_lattice:sec}

\begin{figure}[htb]
\centering
\includegraphics[scale=0.5]{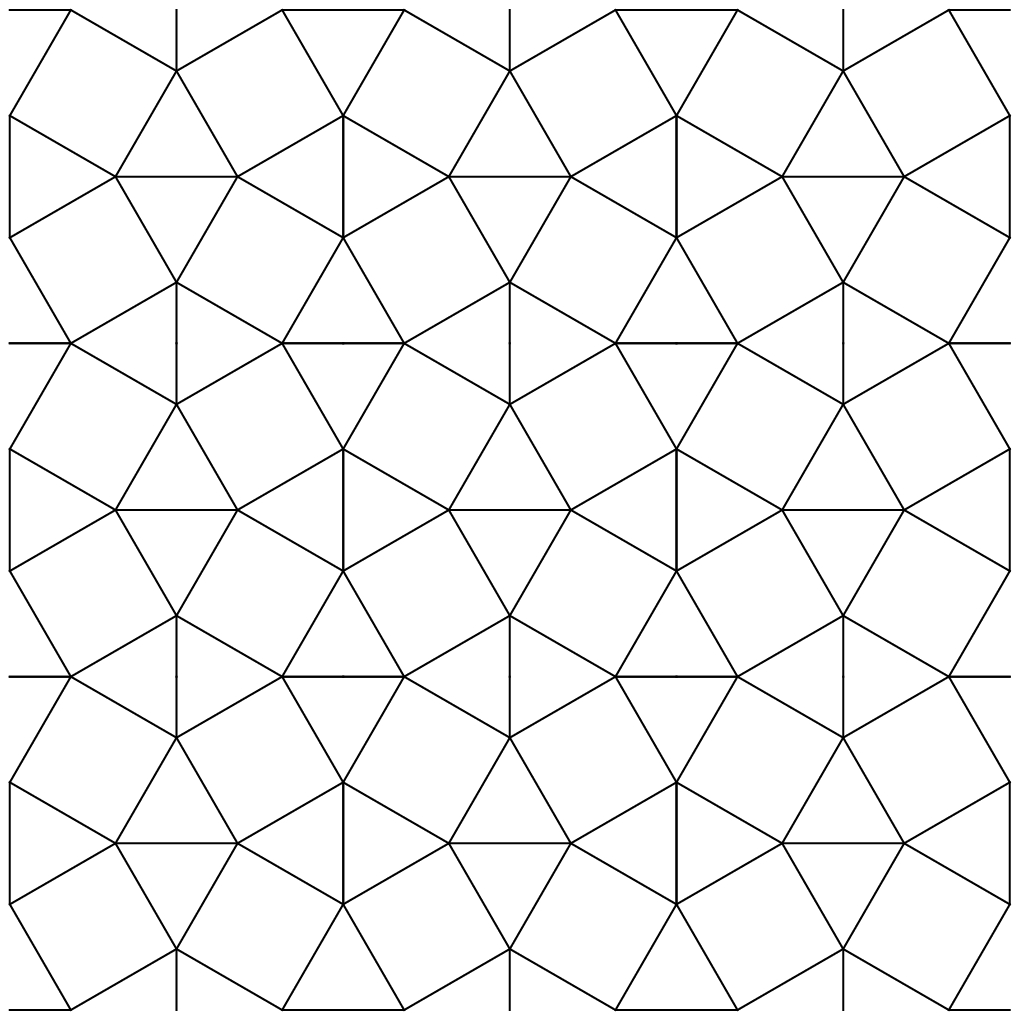}
\caption{\label{snub_lattice:fig}
%(color online)
The lattice is a periodic tiling of the plane by squares and triangles.
}
\end{figure}

\begin{figure}[htb]
\centering
\includegraphics[scale=0.4]{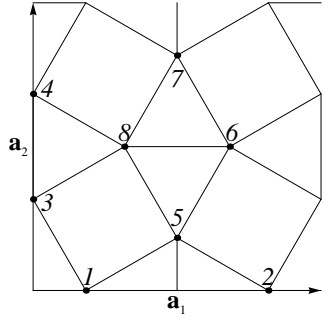}
\caption{\label{snub_unitcell:fig}
%(color online)
The unit cell of the lattice structure shown in figure~\ref{snub_lattice:fig}.
There are eight lattice points in each unit cell, $p=8$.
}
\end{figure}

To demonstrate how efficient our general formalism is for calculating the resistance in a resistor network,
we present results for a more complex lattice.
Consider a network which is a periodic tiling of the plane by squares and triangles
shown in figure~\ref{snub_lattice:fig}.
The unit cell may be chosen as can be seen in figure~\ref{snub_unitcell:fig}.

The Fourier transform of the Laplacian matrix for the lattice shown in figure~\ref{snub_lattice:fig} can easily be found:
\begin{equation}\label{L_snub:eq}
\textbf{L}(\textbf{k}) = \frac{1}{R} \, \left[
\begin{array}{cccccccc}
 -5 & A^* & 1 & B^* & 1 & 0 & B^* & 0 \\[1ex]
 A & -5 & A & C^* & 1 & 0 & B^* & 0 \\[1ex]
 1 & A^* & -5 & 1 & 0 & A^* & 0 & 1 \\[1ex]
 B & C & 1 & -5 & 0 & A^* & 0 & 1 \\[1ex]
 1 & 1 & 0 & 0 & -5 & 1 & B^* & 1 \\[1ex]
 0 & 0 & A & A & 1 & -5 & 1 & 1 \\[1ex]
 B & B & 0 & 0 & B & 1 & -5 & 1 \\[1ex]
 0 & 0 & 1 & 1 & 1 & 1 & 1 & -5
\end{array}
\right],
\end{equation}
where $A = \rme^{\rmi {\bf k} {\bf a}_1}, B = \rme^{\rmi {\bf k} {\bf a}_2}$ and $C = \rme^{\rmi  {\bf k} \left({\bf a}_2 -{\bf a}_1\right)}$.

It is difficult to find analytical result since the integrand in equation~(\ref{Rabmn_int_x1_x2:eq})
is a very complicated function of its variables.
However,  numerically we calculated the resistance $R_{\alpha \beta}(0,0)$ (in units of $R$)
between lattice points $\alpha=1,\dots,8$ and $\beta=1,\dots,8$ that belong to the same unit cell and find
\begin{equation}\label{snub_Rab_num_par:eq}
    R_{\alpha \beta}(0,0) =
\left[
\begin{array}{cccccccc}
 0 & r_4 & r_2 & r_5 & r_2 & r_6 & r_7 & r_3 \\
 r_4 & 0 & r_7 & r_8 & r_2 & r_3 & r_7 & r_6 \\
 r_2 & r_7 & 0 & r_1 & r_3 & r_5 & r_6 & r_2 \\
 r_5 & r_8 & r_1 & 0 & r_6 & r_5 & r_3 & r_2 \\
 r_2 & r_2 & r_3 & r_6 & 0 & r_2 & r_4 & r_2 \\
 r_6 & r_3 & r_5 & r_5 & r_2 & 0 & r_2 & r_1 \\
 r_7 & r_7 & r_6 & r_3 & r_4 & r_2 & 0 & r_2 \\
 r_3 & r_6 & r_2 & r_2 & r_2 & r_1 & r_2 & 0
\end{array}
\right],
\end{equation}
where $r_1 \approx 0.3849$, $r_2 \approx 0.4038$, $r_3 \approx  0.5108$, $r_4 \approx 0.5396$, $r_5 \approx 0.5585$,
$r_6 \approx 0.5647$, $r_7 \approx 0.6230$, $r_8 \approx 0.6631$.
It is interesting to note that the resistance $R_{14}=r_5$ and $R_{16}=r_6$ are almost the same numerically.
However, rigorously they are not the same since they are not related to each other by symmetry.

\subsection{Body centered cubic lattice }\label{bcc_lattice:sec}

\begin{figure}[htb]
\centering
\includegraphics[scale=0.45]{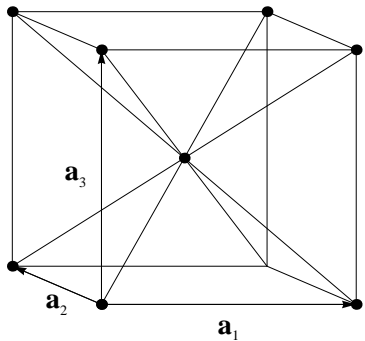}
\caption{\label{bcc_lattice:fig}
%(color online)
The body centered cubic lattice as a periodic tiling of three dimensional space.
Besides the sides of the cube there are resistors between the center of the cube and its corners.
There are two non-equivalent lattice points in the bbc lattice, one is at one of the corner of the cube
and the other is at the center of the cube, ie, $p=2$.
The unit cell vectors are $\textbf{a}_1$, $\textbf{a}_2$ and $\textbf{a}_3$.
All lines represent a resistor with resistance $R$.
}
\end{figure}

Finally, we present a non-trivial example for a three dimensional lattice of a resistor network.
The simple cubic lattice has already been studied in reference~\cite{CsJ_resistor-1:cikk}, 
and Glasser and Boersma calculated the exact values of the resistances for a few cases~\cite{Glasser_Boersma:cikk}.
Consider a more complicated resistor network, namely the body centered cubic (bcc) lattice
shown in figure~\ref{bcc_lattice:fig}.

As we demonstrated above, the Laplacian matrix can be obtained from Ohm's and Kirchhoff's laws and we find that
its Fourier transformation can be written as
\numparts
\begin{eqnarray}
\label{bcc_Lm:eq}
\mathbf{L}(\mathbf{k}) &=&
\frac{1}{R} \left[ \begin{array}{cc}
A-14 & B^*\\
B & -8
\end{array}
\right], \rm{where} \\[1ex]
A &=& 2(\cos {\bf k} {\bf a}_1 + \cos {\bf k} {\bf a}_2 + \cos {\bf k} {\bf a}_3 ),  \\[1ex]
B &=&(1+\rme^{\rmi {\bf k} {\bf a}_1})(1+\rme^{\rmi {\bf k} {\bf a}_2})(1+\rme^{\rmi {\bf k} {\bf a}_3}).
\end{eqnarray}
\endnumparts
Then, the Green's function from equation~(\ref{Green_L:eq}) becomes
\begin{equation}
\mathbf{G}(\mathbf{k})=\frac{1}{\det (\textbf{L})}
\left[\begin{array}{cc}
8 & B^*\\
B & 14-A
\end{array}\right],
\end{equation}
where the determinant of $\textbf{L}(\textbf{k})$ is given by
\begin{eqnarray}
\label{bcc_detL:eq}
\lefteqn{
R \det (\mathbf{L}) = 112-16( \cos{\bf k} {\bf a}_1 + \cos{\bf k} {\bf a}_2 + \cos {\bf k} {\bf a}_3 ) }
\nonumber \\[1ex]
&& -8(1+\cos {\bf k} {\bf a}_1)(1+\cos {\bf k} {\bf a}_2)(1+\cos {\bf k} {\bf a}_3).
\label{eq:bcc determinans}
\end{eqnarray}

The resistance between two arbitrary lattice points can be determined from equation~(\ref{Rabm1md_int_x1_xd:eq}) for $d=3$.
For a few cases we find numerically that the resistances (in units of $R$) are $R_{12}(0,0,0)\approx 0.1945$,
$R_{11}(1,0,0)\approx 0.1481 $, $R_{11}(1,1,0)\approx 0.1651 $, $R_{11}(1,1,1)\approx 0.1717 $ and
$R_{22}(1,0,0)\approx 0.2657 $.
It is interesting to note that the resistance $R_{11}(1,0,0)$ (resistance between the two ends of a side of the cube)
is much less than that in the simple cubic lattice (in this case it is $R/3$, see, eg,~\cite{CsJ_resistor-1:cikk}).
The physical reason for this difference is that comparing with the simple cubic lattice, the body centered cubic lattice 
provides more channels for the current flowing between the two lattice points.

\section{Conclusions}\label{Conclusion:sec}

In this work using the Green's function method we derived a general resistance formula for any infinite lattice structure of resistor networks that is a periodic tiling of space in all dimensions.
Our general resistance formula was applied to several non-trivial resistor networks to demonstrate how versatile our approach is.
For the Kagom\'e and dice lattice we derived explicit expressions for the resistances between two arbitrary lattice points
in terms of the resistances on a triangular lattice of resistors.
Similarly, we showed that there is a direct map between the decorated lattice of resistors and the square lattice of resistors.
We pointed out that a mapping between the lattice structure shown in figures~\ref{4by4_lattice:fig} 
and~\ref{cent_sq_lattice:fig} can exist.
We believe that such a mapping between different lattice structures of resistor networks has a topological explanation. 
In fact we think that there exists a classification of different resistor networks in terms of some classes. 
However, for a deeper understanding of this issue more work needs to be done.
This problem could be a future challenge for physicists and mathematicians.

Tiling of plane is common in the arts and its mathematical description based on group theory is well known in the literature.
Under the title 'tiling' one can find numerous decorative and practical examples of possible tilings on the world-wide web.
Replacing the lines by resistors in such tilings provides a wealth of examples for possible resistor networks not studied 
in the literature.
Here, we presented examples that are well known and relatively simple to find analytical results for.
However, our Green's function method is general and makes it possible to study very complicated lattice structures such as
that discussed in section~\ref{snub_lattice:sec}.

Our work can be extended to study the classical lattice dynamics and the vibrational modes of atoms 
within the framework of the harmonic approximation.
Similarly, description of the electron dynamics governed by the Schr\"odinger equation 
in the tight binding approximation could be another application of the Green's function approach 
outlined here in the case of resistor networks.

\ack

The authors wish to thank L. Glasser, T. Guttmann, L. Lov\'asz, S. Redner and F. Y. Wu, for helpful discussions.
This work was supported by the Marie Curie ITN project NanoCTM (FP7-PEOPLE-ITN-2008-234970)
and the Hungarian Science Foundation OTKA under the contracts No. 75529 and No.~81492.
The European Union and the European Social Fund have provided financial support to the project
under the grant agreement no. TÁMOP 4.2.1./B-09/1/KMR-2010-0003.

\appendix

\section{Relation between the Kagom\'e and the triangular lattice of resistor networks } \label{kagome_3szog:app}

In this appendix we list the results for the resistance $R_{\alpha \beta}(m,n)$ on Kagom\'e lattice
in terms of the resistances $R^\vartriangle(m,n)$ on a triangular lattice:
\numparts
\begin{eqnarray}\label{R_Kagome-3szog:eq}
\fl  R_{11}(m,n) = \frac{R}{9} + \frac{7}{3}\,  R^\vartriangle (m,n)
  -\frac{1}{6}\, \Bigl[R^\vartriangle (m-1,n+1)+ R^\vartriangle (m+1,n-1) \Bigr], \nonumber  \\[2ex]
\fl   R_{12}(m,n)  = \frac{R}{9} + \frac{5}{6}\,  \Bigl[R^\vartriangle (m,n)+R^\vartriangle (m+1,n) \Bigr]
+\frac{1}{6} \Bigl[ R^\vartriangle (m+1,n-1) + R^\vartriangle (m,n+1)\Bigr],  \nonumber  \\[2ex]
\fl  R_{13}(m,n) = \frac{R}{9} + \frac{5}{6}\,  \Bigl[R^\vartriangle (m,n)+R^\vartriangle (m,n+1)\Bigr]
  +\frac{1}{6}\, \Bigl[R^\vartriangle (m+1,n) + R^\vartriangle (m-1,n+1) \Bigr],  \nonumber   \\[2ex]
\fl  R_{22}(m,n) = \frac{R}{9}  + \frac{7}{3}\,  R^\vartriangle (m,n)
  -\frac{1}{6}\, \Bigl[R^\vartriangle (m,n-1) + R^\vartriangle (m,n+1)\Bigr],  \nonumber  \\[2ex]
\fl  R_{23}(m,n) = \frac{R}{9} + \frac{5}{6}\, \Bigl[R^\vartriangle (m,n)+R^\vartriangle (m-1,n+1)\Bigr]
   +\frac{1}{6}\, \Bigl[R^\vartriangle (m-1,n) + R^\vartriangle (m,n+1) \Bigr],  \nonumber   \\[2ex]
\fl  R_{33}(m,n) = \frac{R}{9} + \frac{7}{3}\, R^\vartriangle (m,n)
  -\frac{1}{6}\, \Bigl[R^\vartriangle (m-1,n) + R^\vartriangle (m+1,n)\Bigr]. \nonumber
\end{eqnarray}
\endnumparts

The remaining resistances can be obtained from the symmetry relation
$R_{\alpha \beta} (m,n)= R_{\beta \alpha} (-m,-n)$.

\section{Relation between the dice and the triangular lattice of resistor networks } \label{diced_3szog:app}

Here we list the results for the resistance $R_{\alpha \beta}(m,n)$ on a dice lattice
in terms of the resistances $R^\vartriangle(m,n)$ on a triangular lattice:
\numparts
\begin{eqnarray}\label{R_Dice-3szog:eq}
\fl  R_{11}(m,n) = \frac{3}{2}\, R^\vartriangle (m,n), \nonumber \\[2ex]
\fl  R_{12}(m,n)  = \frac{R}{6} + \frac{1}{2}\, \Bigl[ R^\vartriangle (m,n)+R^\vartriangle (m+1,n)
                    + R^\vartriangle (m,n+1)  \Bigr], \nonumber \\[2ex]
\fl  R_{13}(m,n) = \frac{R}{6} + \frac{1}{2}\, \Bigl[ R^\vartriangle (m+1,n)+R^\vartriangle (m,n+1)
                    + R^\vartriangle (m+1,n+1)  \Bigr], \nonumber  \\[2ex]
\fl  R_{22}(m,n) = \frac{R}{3}\, c_{mn} + \frac{3}{2}\, R^\vartriangle (m,n), \nonumber  \\[2ex]
\fl  R_{23}(m,n) = \frac{R}{3} + \frac{1}{6}\, \Bigl[2 R^\vartriangle (m,n)+ 2 R^\vartriangle (m+1,n) + 2R^\vartriangle (m,n+1) \Bigr. \nonumber \\[1ex]
   \Bigl. + R^\vartriangle (m-1,n+1) + R^\vartriangle (m+1,n-1) + R^\vartriangle (m+1,n+1) \Bigr] , \nonumber  \\[2ex]
\fl  R_{33}(m,n) = R_{22}(m,n), \nonumber
\end{eqnarray}
\endnumparts
where $c_{mn} =0 $, if $m$ and $n$ equal to zero, otherwise it equals to $1$.
The remaining resistances can be obtained from the symmetry relation
$R_{\alpha \beta} (m,n)= R_{\beta \alpha} (-m,-n)$.

\section{Relation between the decorated lattice and the square lattice of resistor networks } \label{Lieb_4szog:app}

Here we list the results for the resistance $R_{\alpha \beta}(m,n)$ on a decorated lattice
in terms of the resistances $R^\square (m,n)$ on a square lattice:
\numparts
\begin{eqnarray}\label{R_Lieb-4szog:eq}
\fl  R_{11}(m,n) = 2\, R^\square (m,n),  \nonumber  \\[2ex]
\fl  R_{12}(m,n) = \frac{R}{4} +  R^\square (m,n)+R^\square (m+1,n) ,  \nonumber  \\[2ex]
\fl  R_{13}(m,n) =  \frac{R}{4} + R^\square (m,n)+R^\square (m,n+1) ,  \nonumber \\[2ex]
\fl  R_{22}(m,n) = \frac{R}{2} + 3\, R^\square (m,n)  - \frac{1}{2}\, \Bigl[R^\square (m,n-1)+ R^\square(m,n+1)\Bigr],  \nonumber  \\[2ex]
\fl  R_{23}(m,n) =  \frac{R}{2} + \frac{1}{2}\, \Bigl[ R^\square (m,n)  + R^\square (m-1,n)
                  + R^\square(m-1,n+1)+  R^\square (m,n+1)\Bigr] ,  \nonumber \\[2ex]
\fl  R_{33}(m,n) = \frac{R}{2} + 3\, R^\square (m,n) - \frac{1}{2}\, \Bigl[R^\square (m-1,n)+ R^\square(m+1,n)\Bigr].  \nonumber
\end{eqnarray}
\endnumparts
The remaining resistances can be obtained from the symmetry relation
$R_{\alpha \beta} (m,n)= R_{\beta \alpha} (-m,-n)$.

\section*{References}

%\bibliography{c:/cserti/AD/Graphene/BibTeX-graphene/graphene,c:/cserti/AD/BibTeX/cikkek,c:/cserti/szem/publ/sajatcikkek_CsJ}
%\bibliography{/home/cserti/AD/Graphene/BibTeX-graphene/graphene,/home/cserti/AD/BibTeX/cikkek,/home/cserti/szem/publ/sajatcikkek_CsJ}

%\bibliographystyle{iopart-num}
%\bibliographystyle{prsty}
%\bibliographystyle{/usr/lib/texmf/texmf/tex/revtex/osa}

%\end{document}

\providecommand{\newblock}{}

\end{document}